\documentclass[11pt, a4paper]{article}
\usepackage{jheppub}
\usepackage{color}
\usepackage{graphicx}
\usepackage{caption}
\usepackage{epsfig}
\usepackage{dcolumn}
\usepackage{bm}
\usepackage{float}
\usepackage{hyperref}
\usepackage{slashed}
\usepackage{ifpdf}
\newcommand{\beq}{\begin{equation}}
\newcommand{\eeq}{\end{equation}}
\newcommand{\bea}{\begin{eqnarray}\begin{aligned}}
\newcommand{\eea}{\end{aligned}\end{eqnarray}}
\newcommand{\Veff}{V_{\text{eff}}}
\newcommand{\mttilde}{m_{\tilde{t}}}
\newcommand{\leff}{\lambda_{\text{eff}}}
\newcommand{\lnbar}{\overline{\ln}}
\newcommand{\gev}{\;\text{GeV}}
\newcommand{\tev}{\;\text{TeV}}
\newcommand{\mev}{\;\text{MeV}}
\newcommand{\wtilde}{\widetilde{W}}

\title{Leading Two-loop corrections to the mass of Higgs boson in the High scale Dirac gaugino supersymmetry}
\author{Di Liu}
\affiliation{Center for Cosmology and Particle Physics, Department of Physics, New York University, New York, NY 10003, USA}
\emailAdd{di.liu@desy.de}

\abstract{Precision measurements of the Higgs mass have become a powerful constraint on models of physics beyond the standard model. We revisit supersymmetric models with Dirac gauginos and study the contributions to the Higgs mass. We calculate the leading two-loop corrections to the SM-like Higgs mass by constructing a series of EFTs and iteratively integrating out heavy particles. We then apply these calculations to a variety of scenarios, including a simple Dirac gluino, and split Dirac models of supersymmetry. We present the detailed formulae for threshold corrections and compare with previous results, where available. In general, the contributions are small, but the additional precision allows us to make more concrete statements about the relevant scales in Dirac SUSY models.}

\keywords{Higgs boson, Two-loop, Dirac gaugino}

\arxivnumber{1912.06168}

\begin{document}

\maketitle
\flushbottom

\section{Introduction}
The discovery of the Higgs Boson was a triumph of the standard model, providing the final piece after a wait of a half-century. For physics beyond the standard model, it is a critical piece, as the hierarchy problem is essentially a question about the Higgs mass. The increasingly precise measurements of its properties so far point to a very standard model-like Higgs boson, leaving questions as to where the new physics lies.

In supersymmetric theories, there is a close relationship between the masses of the new supersymmetric partners and the Higgs mass. Top-partners, or stops, in particular, are important. For minimal SUSY the Higgs mass of 125\gev\, points to stops typically in the 10\tev\, range. Such heavy stops typically correct the soft mass-squared of the Higgs fields, leading to percent (or worse) tuning to achieve a physical mass of 125 GeV. This is exacerbated if the stop masses are generated at a high scale, enhancing the Higgs mass corrections by large logarithms.

The consequence of this is that people have pushed in new directions to understand how the Higgs mass can be so much smaller than the scale of new physics. One path is that one can search for models where the large logarithms are absent, and the Higgs mass is naturally (or more naturally) at its observed scale. Alternatively, one can assume that the large tuning is present but solved through something like anthropic selection. Interestingly, Dirac gauginos play roles in both these possibilities.

In the MSSM, gauginos are typically taken to be Majorana. But with Dirac gauginos, the radiative corrections to scalar masses are finite, allowing one to raise gaugino and squark masses without introducing large logs. Simultaneously, the different symmetry properties of Dirac gauginos allows one to consider new ``split'' scenarios~ \cite{Fox:2005yp, Fox:2014moa} where gauginos appear at a very high energy scale.

Because the Higgs mass is so well measured, it can have significant implications for the scales of new physics in SUSY models. In particular, it is well known that two-loop corrections in the MSSM play an important role in setting the scale of squarks and gluinos \cite{Heinemeyer:1998kz, Papucci:2011wy, Buckley:2016tbs}. 
Here, we aim to pursue this line of thinking to examine the role of two-loop effects in Dirac supersymmetry. The reasons for this are clear: given the precision of the Higgs mass, we want a similar level of accuracy in our calculation. Secondly, it has been argued that the two-loop effects in Dirac models can be large \cite{Diessner:2015yna}, although later studies have argued that the on-shell consequences are small \cite{Braathen:2016mmb}, warranting a confirmation. Finally, no one has extended the two loop thresholds to study heavy SUSY, where the experimental data appears to be pushing us.

In this paper, we will study the two-loop thresholds to Dirac SUSY in a variety of regimes. We will begin by developing our formalism and then applying it to specific scenarios. We find, agreeing with \cite{Braathen:2016mmb} that the contributions are small for low-scale Dirac gluinos. We similarly find the effects are small for high scale gauginos, allowing us to robustly interpret the predictions of previously proposed split-Dirac scenarios.

\subsection{The Higgs Boson Mass in Supersymmetry}\label{higgsmassintro}
In low-scale supersymmetry ($m_S\simeq $ 1TeV), quantum corrections to the SM-like Higgs boson mass $m_h$ can be calculated through the ``fixed-order'' method, i.e., using a diagrammatic approach \cite{Haber:1990aw, Okada:1990vk, Ellis:1991zd, Heinemeyer:1998jw, Heinemeyer:1998np, Heinemeyer:1998kz, Martin:2004kr}, evaluating the full set of Feynman diagrams for the self-energy  of the Higgs boson. Since the Higgs mass is evaluated at the low energies, the external momenta of the Feynman diagrams can be set to zero to simplify the calculation. Equivalently, one can use the effective potential approach \cite{Zhang:1998bm, Espinosa:2000df, Diessner:2015yna, Braathen:2016mmb, Hempfling:1993qq} which extracts $m_h$ from the derivatives of the effective potential. Since the tree-level mass for the Higgs boson in the MSSM is bounded from above by $M_Z$, the radiative corrections are necessarily sizable to maintain viability. Studies have found that two-loop effects are important in understanding the implications for SUSY spectra \cite{Braathen:2019zoh, Bagnaschi:2019esc, Hollik:2015ema, Hahn:2014qla, Draper:2013oza, Degrassi:2009yq, Buckley:2016tbs}.

\begin{figure}[h!]
\centering
\includegraphics[width=8.0cm, height=2.0cm]{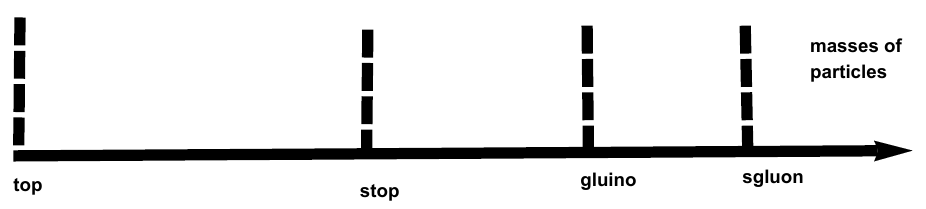}
\caption{\small{Hierachy of the mass spectrum}}
\label{fig:masshier}
\end{figure}

With the experimental data at hand, it appears there is a sizable separation between the electroweak scale and the SUSY scale, i.e., $\mttilde \gg m_t$, if SUSY is present in nature at all. In this case, or when there exists hierarchical split in the SUSY mass spectrum\cite{Giudice:2011cg}, the ``fixed-order'' approach may become inadequate because of the large logarithms between mass thresholds~\cite{Bagnaschi:2014rsa, Draper:2016pys,Hahn:2009zz}. In this case, the Higgs mass determination needs to be realized in an effective field theory (EFT) approach: heavy particles are integrated out at the scale $\tilde{m}_i$, where they only induce threshold corrections; then the corresponding renormalization group equations (RGEs) are used to evolve the renormalized couplings from one mass threshold to another. The mass of the Higgs boson is eventually determined at the EWSB scale. In this procedure, the threshold corrections are free of large ratio of scales and the low scale EFT parameters do not blow up. 
In this paper, we study the pole mass of Higgs boson of the Dirac-gaugino model\cite{Fox:2002bu} and will pursue the  EFT approach due to the hierarchy mass spectrum predicted by this model as it is shown in FIG.~(\ref{fig:masshier}). \\

In pursuing the EFT approach, we will define the relevant energy thresholds and intervals as,
\begin{itemize}
\item For the energy scale $M_{O_S}<Q<M_{UV}$, the Higgs quartic coupling $\lambda$ follows the SUSY relation: $\lambda=\frac{1}{4}(g_Y^2+g_2^2)\cos^2(2\beta)$.

\item At $Q=M_{O_S}$, we integrate out the sgluon, $\lambda$ gets the first threshold correction form the two loop diagram involving stop-sgluon-stop.

\item For $M_3<Q<M_{O_S}$, $\lambda$ runs down through its $\beta$-function, from $\lambda(M_{O_S})$ to $\lambda(M_3)$

\item At $Q=M_3$, gluino is integrated out, and $\lambda$ gets the second threshold correction from top-stop-gluino diagram. 

\item For $m_{\tilde{t}}<Q<M_3$, $\lambda(M_3)$ runs down again and produces $\lambda(m_{\tilde{t}})$.

\item At $Q=m_{\tilde{t}}$, $\lambda$ gets the third threshold correction from all one-loop SM superpartners involved diagrams and two loop stop-stop, stop-gluon-stop diagrams.

\item For $Q<m_{\tilde{t}}$, $\lambda$ becomes an SM parameter, and runs down with SM RGEs at $Q=m_t$ to compute the Higgs pole mass at the leading 2-loop order.
\end{itemize}

The paper is organized as the following: after setting up the important elements of our Dirac-gaugino model in Sec.~\ref{setup}, we clarify the general method of our calculation in Sec.~\ref{principle}. In Sec.~\ref{sg} to \ref{st}, we go through the details of constructing EFTs in such a way that heavy particles are iteratively integrated out; several complementary expression of the threshold corrections are given in Sec.~\ref{appendix}. In Sec.~\ref{num}, the numerical impacts to the Higgs mass from 1) Dirac gluino model 2) Pure Dirac model 3) Hypercharge impure model are discussed. We give a conclusion in Sec.~\ref{conclusion}. 

\section{Dirac Gauginos}\label{setup}
SUSY models with Dirac gauginos have two key elements beyond usual MSSM models. First, SUSY must be broken by a $D$-component vev of a hidden sector $U(1)'$, whose field strength we denote by $W'_\alpha$. Second, the model must be extended by adding adjoint chiral superfields \cite{Fox:2002bu} to form as Dirac partners to the MSSM gauginos. We shall denote collectively $\mathbf{\Sigma^a}=\Sigma^a+\sqrt{2}\theta\chi_\Sigma^a+\cdots$, where the lowest-order component $\Sigma^a$ is a complex scalar. 
\begin{equation}
-\mathcal{L}_{\text{supersoft}}=\int d^2\theta\sqrt{2}\frac{W'\cdot W^a\mathbf{\Sigma^a}}{M}+ \text{h.c.}\supset m_D\tilde{A}^a\chi_\Sigma^a+\sqrt{2}m_D \Sigma^a D^a +\text{h.c.}\;,
\end{equation}

where $m_D=D'/M$, $W^a$ represent the field-strength superfields. $\tilde{A}^a$ and $D^a$ are the fermion component (gaugino) and the $D$-term of $W^a$ respectively. This operator is \textit{supersoft} in that the radiative corrections it generates for other soft masses are finite, in contrast with \textit{e.g.}, a Majorana gaugino mass. 
We shall denote the adjoint multiplet for $SU(2)_L$ as a triplet $\mathbf{T^a}=T^a+\sqrt{2}\theta\chi_T^a+\cdots$, and the  $SU(3)_c$ as a octect $\mathbf{O^a}=O^a+\sqrt{2}\theta\chi_O^a+\cdots$. Including this operator and integrating out the auxiliary field $D^a$, the Lagrangian contains new interactions involving the gluinos and sgluons. We express sgluon field in terms of its real and imaginary components: $O^a=(O^a_S+i O^a_A)/\sqrt{2}$, 
\bea\label{lgaugino}
-\mathcal{L}_{\text{g}}&\supset M_3 \overline{\tilde{g}_D^{a}}\tilde{g}_D^{a}+\frac{1}{2}M_{O_S}^2O_S^aO_S^a+\frac{1}{2}m_{O_A}^2 O_A^aO_A^a+2g_sT^{aj}_{i}M_3(\tilde{Q}_{3}^{\dagger i}\tilde{Q}_{3j}-\tilde{u}_3^{\dagger i}\tilde{u}_{3j})O_S^a
\eea
where $ M_3$ is the mass of the Dirac-gluino. The mass of the scalar $O_S$ and the pseudoscalar $O_A$ are related by the mass splitting formula $ M_{O_S}^2=4M_3^2+m_{O_A}^2$, where $m_{O_A}^2$ is an additional soft-breaking mass to allow for more general SUSY breaking scenarios. Our extended SUSY model described by e.q.~(\ref{lgaugino}) has the following properties: 
$i)$ The Dirac-gaugino $\tilde{g}_D$ consists of the MSSM gluino $\tilde{g}^a$ and the extra Weyl fermion octect $\chi_{O}^a$.
$ ii)$ There are new trilinear terms between the ``sgluon'' $O^a$ and stops which have no analog in the MSSM but play an important role in the model.

Besides E.q.~(\ref{lgaugino}), the dominant interactions involved in the Higgs quartic calculation also include, 
\begin{equation}\label{lmssm}
-\mathcal{L}\supset \frac{\lambda_D}{2}(H^\dagger H)^2+ m_Q^2 \tilde{Q}_3^\dagger \tilde{Q}_3+m_U^2\tilde{u}_3^\dagger \tilde{u}_3+\xi_{\tilde{t}}(|H\tilde{Q}_3|^2+|H\tilde{u}_3|^2)+\zeta_{\tilde{t}}[(\tilde{Q}_3^\dagger \tilde{Q}_3)^2+(\tilde{u}_3^\dagger \tilde{u}_3)^2]\;,
\end{equation}
where $m_Q^2$ and $m_U^2$ are the soft SUSY-breaking mass terms for the stops. Regarded as the full theory, the tree level Lagrangian E.q.~(\ref{lgaugino}), (\ref{lmssm}) has three characteristic mass scales in the following hierarchy: $M_{O_s}>M_3>m_Q, m_U$ and other dimensionless model parameters satisfy the relation provided by SUSY,
\begin{equation}\label{susylim}
\lambda_D= \frac14\left(g_2^2+\frac35\, g_1^{2}\right)\cos^22\beta~,\quad  g_t=y_t\sin\beta\;,\quad \xi_{\tilde{t}}=g_t^2\;,\quad \zeta_{\tilde{t}}=\frac{g_s^2}{6}, 
\end{equation}
where, $g_1$ is the $SU(5)$ normalized hypercharge gauge coupling, $g_2$ is the weak gauge coupling and $g_t$ is the top Yukawa coupling.  As is well known, in the presence of Dirac gaugino masses, integrating out massive adjoint scalars eliminates the tree-level $D$-term quartics~\cite{Fox:2002bu}. To focus for the moment on the corrections arising from the gluino, we limit ourselves to considering a Dirac gluino alone - no adjoint superfields for $SU(2)_L\times U(1)_Y$ sectors are introduced.

\section{Match and Run in the mass spectrum}\label{matchrun}
With this model in mind, we will use EFT and RG techniques to compute the leading two-loop radiative corrections to the SM-like Higgs quartic $\lambda$ in the scenarios where the soft Dirac-gluino mass is much larger than the soft stop mass. This technique is both efficient and accurate in calculating the couplings of the SM EFT in the sense that radiative corrections containing ``large logarithms'' are resolved by the solutions of RGEs and the loop corrections from heavy particles in the UV theories only appear in the threshold matching conditions for the IR EFTs.

\subsection{General procedure of the threshold matching}\label{principle}
Threshold Higgs quartic matching is accomplished by comparing  $H H \rightarrow H H$ scattering amplitudes $\mathcal{M}(p_1, \cdots , p_4)$ calculated in both UV and IR theories on the boundary. Since the value of $p_i$ has nothing to do with the UV dynamics\cite{Skiba:2010xn},  
we will make an opportunistic choice of zero external momenta so that the quantum corrections $\Delta\mathcal{M}^{\text{loop}}$ are able to be depicted by derivatives of the vacuum bubble diagrams\cite{Quiros:1999jp, Coleman:1973jx, Jackiw:1974cv, Draper:2016pys}, aka Coleman-Weinberg potential $V_{\text{cw}}(H)$,
\begin{equation}\label{vder}
 \Delta\mathcal{M}^{\text{loop}}(\textbf{0})=\lambda_{\text{cw}}= \frac{1}{2}\frac{\partial^4V_{\text{cw}}(H)}{\partial^2 H^\dagger\partial^2 H}\bigg\vert_{H\rightarrow 0}\;.
\end{equation}
Including the tree-level terms in the renormalized Lagrangian, one can further define the effective potential for the Higgs boson $\Veff=-m_R^2 H^\dagger H+\frac{\lambda_R}{2} (H^\dagger H)^2+V_{\text{cw}}$. As a consequence,  a sum of one-particle irreducible 4-point function at vanishing external momenta is represented by,
 \begin{equation}\label{lambdacw}
 \leff=\lambda_{\text{R}}(Q)+\lambda_{\text{cw}}(Q)=\mathcal{M}(\textbf{0})\;,
 \end{equation}
where the value of $\lambda_R$ and $\lambda_{\text{cw}}$ depends on the renormalization scale while $\leff$ keeps invariant when $Q$ changes. However, being evaluated from the fixed order computation, the result of $\mathcal{M}(\textbf{0})$ will become unreliable if the massive particles ($\tilde{m}_i$) involved in the loop calculation are much heavier than the scale of EWSB; perturbative method breaks down because of the large scale ratio. For this reason, we do not use the actual value of $\leff$ to do any direct prediction, rather, will utilize its property of rescale invariance to match the Higgs quartic threshold correction, $\lambda_{\text{thr}}=\lambda_R^{\text{IR}}(\tilde{m}_i)-\lambda_R^{\text{UV}}(\tilde{m}_i)$.  
Consider the Callan-Symanzik equation applied to $\leff$ at two-loop order,
\bea\label{2lcs}
\frac{\partial\leff^{(2)}}{\partial t}+\beta_{x_i}^{(1)}\frac{\partial\leff^{(1)}}{\partial x_i}+\beta_{x_i}^{(2)}\frac{\partial\leff^{(0)}}{\partial x_i}=0\;,
\eea
where $t=\ln Q$, $\beta_{x_i}=\partial_t x_i$ and $x_i$ represent all the renormalized parameters involved in the calculation like $g_t, g_s$ and $\ln m_i$. At two-loop level $\lambda_{\text{thr}}^{(2)}$ can be then determined at the boundary, 
\bea\label{2Lmatch}
\lambda_{\text{thr}}^{(2)}=\left[\lambda_{\text{cw, UV}}^{(2)}-\lambda^{(2)}_{\text{cw, IR}}+\frac{\partial\leff^{(1)}}{\partial x_i}\Delta x_i^{(1)}+\frac{\partial\leff^{(0)}}{\partial x_i}\Delta x_i^{(2)}\right]_{Q=\tilde{m}}\;, 
\eea
where $\Delta x_i=x_{i, \text{UV}}-x_{i, \text{IR}}$ refers to the threshold correction to the model parameters.\footnote{In case of $\frac{\partial\leff}{\partial x_i}$ becomes discontinuous when passing across the boundary, the replacement should be implemented,
\beq
\frac{\partial\leff^{(m)}}{\partial x_i}\Delta x_i^{(n)}\longrightarrow \left(\frac{\partial\leff^{(m)}}{\partial x_i}x_i^{(n)}\bigg\vert_{\text{UV}}-\frac{\partial\leff^{(m)}}{\partial x_i} x_i^{(n)}\bigg\vert_{\text{IR}}\right)
\eeq} The terms proportional to $\Delta x_i^{(i)}$ play the similar role of $\Delta \lambda^{2\ell, \text{shift}}$ in ref. \cite{Bagnaschi:2014rsa, Bagnaschi:2017xid}, which cancel out the apparent IR divergences induced from large ratios of mass scales. Between mass thresholds, $\lambda_R$ runs according to RGEs and at the electroweak scale it relates to Higgs pole mass via, 
\beq\label{mhpole}
[m^{\text{pole}}_h]^2=2[\lambda_R(m_t)+\delta\lambda(m_t)]v_h^2
\eeq
where $v_h=174$\gev, and $\delta\lambda(m_t)\simeq 0.06$ are the SM threshold corrections extracted from the SusyHD~\cite{Vega:2015fna}. We remark that this procedure is accurate only if $\tilde{m}_i\gg v_h$, and therefore terms suppressed by powers of $v_h/\tilde{m}_i$ can be ignored.

In our model, the two-loop effective potential for Higgs boson $\Veff^{(2)}$ depends on $H$ throughout the squared top or stop masses,
\beq
m_t^2=g_t^2|H|^2\;,\quad m_{\tilde{t}_{1,2}}^2=\frac12\left[(m_Q^2+m_U^2+2 g_t^2|H|^2)\pm\sqrt{(m_Q^2-m_U^2)^2+4|X_t|^2}\;\right]\;.
\eeq
In this section, we adopt the so-called ``gaugeless limit'' in our calculation for simplicity, i.e. we neglect the loop corrections contributed by electroweak gauge coupling $g_2$ and $g'$. Moreover, since the quantum corrections from the gluino and sgluon are independent of the quark chirality and are irrelevant to stop mixing we will take $m_Q^2=m_U^2=\mttilde$ and $X_t\approx 0$  through out the work and denote the degenerate stop mass by $\mttilde$. In evaluating the radiative corrections in various frameworks, i.e. SUSY or non-SUSY theories, care must be taken while adopting regularization and renormalization schemes. Working in  non-SUSY models like the SM,  dimensional regularization (DREG) with the $\overline{\text{MS}}$ renormalization scheme is preferred. Although Lorentz symmetric, DREG is well known for its spurious violation of supersymmetry. This is because DREG introduces a continuous spacetime dimensions $ d=4-2\epsilon $ for both momenta and vectors, which leads to a mismatch between the numbers of gauge boson degrees of freedom and the gaugino degrees of freedom off-shell\cite{Martin:2001vx, Martin:1997ns, Martin:1993yx}. In order to obtain a supersymmetric regularization scheme, one can complement the missing dimensions of the gauge boson with an extra $2\epsilon$-dimensional components. Such extra components transform like scalars in the adjoint representations of the gauge group, and are known as epsilon-scalars. A regularization and renormalization scheme based on epsilon-scalars and the modified minimal subtraction, which preserves both Lorentz symmetry and supersymmetry, is called dimensional reduction (DRED) and $\overline{\text{DR}}$ respectively\cite{Siegel:1979wq, Capper:1979ns}. In table~\ref{schemedim}, we list the dimensions of momenta ($p_\mu$), gauge boson vector ($V_\mu^a$), Weyl fermions ($\lambda_\alpha^a, \lambda_{\dot{\alpha}}^a)$ and the epsilon-scalar ($\epsilon^a$) in both $\overline{\text{MS}}$ and $\overline{\text{DR}}$ renormalization schemes, 
\begin{table}[h]
\caption{Comparison of Schemes}
 \begin{center}
  \begin{tabular}{ | c | c | c | c |c| c|}
  \hline
 Dimension  & $p_\mu$ & $V_\mu^a$& $\epsilon^a$& $\lambda_\alpha^a$& $\lambda_{\dot{\alpha}}^a$ \\
   \hline
   $\overline{{\text{MS}}}$& $4-2\epsilon$ & $4-2\epsilon$ & --& 2&2 \\
   \hline
   $\overline{{\text{DR}}}$&$4-2\epsilon$&$4-2\epsilon$&$2\epsilon$&2&2\\
   \hline
  \end{tabular}\label{schemedim}
  \end{center}
\end{table}

where the upper indices ``a'' indicates that the variables are in the adjoint representation of the corresponding gauge group.
From table~\ref{schemedim}, we see that the two renormalization schemes are differed by an additional scalar $\epsilon^a$. Therefore, the difference of the $\overline{{\text{DR}}}$ and $\overline{{\text{MS}}}$ effective potential can be organized from the extra epsilon-scalars with multiplicity $2\epsilon$ in the $\overline{{\text{MS}}}$ scheme.   
\begin{equation}
	V^{(2)}_{\overline{{\text{DR}}}}=V^{(2)}_{\overline{{\text{MS}}}}+V^{(2)}_{\overline{\text{MS}},\epsilon}
\end{equation}
where $V^{(2)}_{\overline{\text{MS}},\epsilon}$ denotes the new contribution from the two-loop vacuum Feynman diagrams when the vector lines are turned into epsilon-scalar lines. The loop integral of epsilon-scalars is discussed in \cite{Capper:1979ns}, and a complete set of $V^{(2)}_{\overline{\text{MS}},\epsilon}$ is given in \cite{Martin:2001vx}. 

In this section, we provide the generic expression of $\Veff^{(2)}$ for the top and stop which are charged under the $SU(N_c)$ gauge group here and will discuss details of the leading two-loop correction in the next few sections and in the Sec.~\ref{fullhiggsqtc},
\bea\label{veffs}
\frac{\Veff^{(2)}}{(N_c^2-1)\kappa^2 g^2}&=\frac{n_F}{4} F_{SS}(\tilde{t},\tilde{t})+M_D^2 n_F F_{SSS}(\tilde{t},\tilde{t}, \Sigma_S)+n_F F_{FFS}(\tilde{A}, t,\tilde{t})\\
&+\frac{n_F}{4}\left[F_{FFV}(t, t, 0)-m_t^2F_{\overline{FF}V}(t, t, 0)\right]+\frac{n_F}{4} F_{SSV}(\tilde{t}, \tilde{t}, 0)
\eea
where $\kappa=(16\pi^2)^{-1}$ and $n_F$ counts the number of fundamental $SU(N_c)$ Weyl fermion: $n_F=2$ for $SU(3)_c$ and $n_F=3$ for $SU(2)_L$. $F_{SS}$, $F_{SSS}$ etc. are loop functions and their explicit expression calculated in both $\overline{\text{DR}}$ and $\overline{\text{MS}}$ schemes are given in \cite{Martin:2001vx, Martin:2002iu}. In the high energy scale, top quarks do not contribute to threshold corrections but caution should be taken when switching from one renormalizion scheme to another. As we will see in the Sec.~\ref{intgluino}, such an issue has to be addressed when the gluino is integrated out.  

\subsection{Integrating out sgluon}\label{sg}
The sgluon is the heaviest particle in the full theory. Integrating out this particle at tree level through the new trilinear interaction results a negative contribution to stop-quartic interaction at low energy scale ($Q\leq M_{O_S}$).
\begin{figure}[h!]
\includegraphics[width=7.5cm, height=3cm]{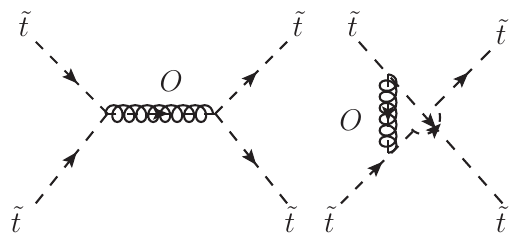}\qquad
\includegraphics[width=3cm, height=3cm]{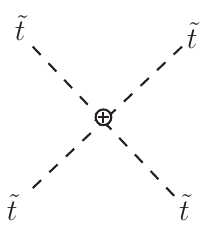}
\caption{Effective stop quartic coupling induced from integrating out sgluon.}
\end{figure}

As a consequence, the original stop self-coupling in Eq.~(\ref{susylim}) reduces to an effective parameter in the IR regime and contributes to $\lambda_{\text{thr}}$ on the sgluon mass threshold. Thus the dominant effective potential contributions relating to threshold matching at $Q=M_{O_S}$ are $V_{\tilde{t}O^a\tilde{t}}^{(2)}(\text{High})$,  $V^{(2)}_{\tilde{t}\tilde{t}}(\text{High})$ , and   $V^{(2)}_{\tilde{t}\tilde{t}}(\text{Low})$\\
\begin{figure}[h!]
\centering
\includegraphics[width=0.40\textwidth]{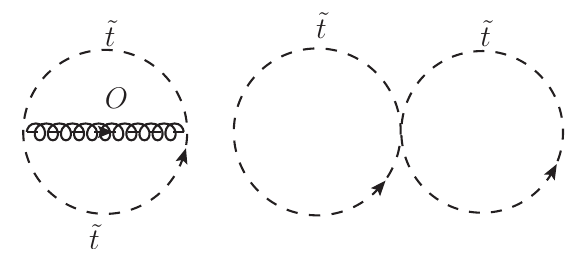}\qquad
\includegraphics[width=0.30\textwidth]{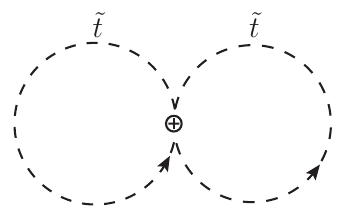}\qquad
\caption{$V_{\tilde{t}O^a\tilde{t}}^{(2)}(\text{High})$,\quad  $V^{(2)}_{\tilde{t}\tilde{t}}(\text{High})$ ,\quad  $V^{(2)}_{\tilde{t}\tilde{t}}(\text{Low})$}\label{fig:stostnstst}
\end{figure}
\\
In our model, the squarks are much lighter than sgluons but have a lower bound $[\mttilde^{\text{pole}}]^2\geq 0.047 M_3^2$ due to sizable radiative corrections from the gluino and sgluon\cite{Fox:2002bu}. In units of $u_0=8\kappa^2g_3^2g_t^4$, the leading order two-loop correction to $\lambda^{(2)}_{\text{cw}}$ from E.q. (\ref{lambdacw}) and (\ref{vder}),
\begin{align}
&u_0^{-1}\lambda^{(2)}_{\tilde{t}O^a\tilde{t}, \text{cw}}=4R\bigg[-\frac{1}{r_s}-\frac{\pi^2}{3}-4\ln r_s-\ln^2 r_s+2r_s(2-\pi^2-10\ln r_s-3\ln^2 r_s)+\mathcal{O}(r_s^2)\bigg]\nonumber\\ 
&u_0^{-1}\lambda^{(2)}_{\tilde{t}\tilde{t}, \text{cw}}(\text{High})=-1+\ln r_s(1+\ln r_s)\;,\quad \lambda^{(2)}_{\tilde{t}\tilde{t}, \text{cw}}(\text{Low})=(1-4R)\lambda^{(2)}_{\tilde{t}\tilde{t}, \text{cw}}(\text{High})\;,
\end{align}
where $r_s=\mttilde^2/M_{O_S}^2$ and  $R=M_3^2/M_{O_S}^2\leq 1/4 $. In the so-called \textit{supersoft} scenario, where soft SUSY-breaking masses vanish, both $\lambda^{(2)}_{\tilde{t}O^a\tilde{t}, \text{cw}}$ and $\lambda^{(2)}_{\tilde{t}\tilde{t}, \text{cw}}$ blow up due to the large $r_s^{-1} $ and $\ln r_s$ in their expressions. However, these apparent IR divergences are cancelled out when matching $\lambda_{\text{eff}}$ on the mass threshold and will not show up in $\lambda_{\text{thr}}^{(2)}$. To be specific, model parameters $\xi_{\tilde{t}}$ and $m_{\tilde{t}}$ receive one-loop quantum corrections
$\xi_{\tilde{t}}^{\text{IR}}= g_t^2(1-\Delta\xi_{\tilde{t}})$ and $(m^2_{\tilde{t}})^{\text{IR}}= m^2_{\tilde{t}}-\Delta m_{\tilde{t}}^{2}$ 
with $\Delta m_{\tilde{t}}^{2}\simeq 16\kappa g_3^2 M_3^2/3$ and $\Delta \xi_{\tilde{t}}\simeq 8\kappa g_3^2 R$, and they will contribute to the two-loop Higgs quartic coupling through the one-loop diagram with stops i.e. $\lambda^{1\ell}_{\tilde{t}}=6\kappa  \xi_{\tilde{t}}^2\lnbar m_{\tilde{t}}^2$ and thus one has,
\begin{equation}\label{dlsg}
 u_0^{-1}\lambda_{\text{thr}}^{(2)}(M_{O_S})\simeq -4R\bigg[\frac{\pi^2}{3}-\frac 12+\left(2\pi^2 -\frac{11}{2}+14\ln r_s\right)r_s+\dots\bigg]\;, \quad (r_s\ll 1) \end{equation}  
the expression remains finite for $r_s\rightarrow 0$ and 
therefore 
the threshold correction to the 4-point Higgs correlation function is free from IR divergence. 
As a renormalized parameter, $\lambda_R=\lambda_D$ is determined from $g_{1, 2}$ as shown in E.q.~(\ref{susylim}) in the UV theory 
and starts to deviate from that expression due to the threshold corrections when $Q<M_{O_S}$. In general we have, 
\begin{equation}
\lambda_R(Q)=\lambda_D(Q)- \int_{M_{O_S}}^Q\sum_i \Delta \beta_\lambda (Q')\Theta(\tilde{m}_i-Q') d\ln Q'+\sum_i \lambda_{\text{thr}}(\tilde{m}_i)\Theta(\tilde{m}_i-Q)
\end{equation}
where $\Theta (x)$ is the Heaviside function. 
On one hand, the RGEs of $\lambda$ get corrections at each threshold, $\beta_\lambda^{\rm thr}(\tilde{m}_i)\equiv\beta_\lambda(\tilde{m}_i^+)-\beta_\lambda(\tilde{m}_i^-)$. On the other hand, the top sector contribution to the Higgs quartic $\beta$-function follows $\beta_\lambda^{\rm top}=12\kappa(\xi_{\tilde{t}}^2-g_t^4)$, the threshold corrections of $\xi_{\tilde{t}}$ and $g_t$ also contribute to the RG running of $\lambda$. Therefore, at the leading order, we take $\Delta \beta_\lambda(Q)\simeq \beta_\lambda^{\rm thr}(\tilde{m}_i)+ 24\kappa g_t^4[\Delta \xi_{\tilde{t}}(Q)- 2\Delta g_t(Q)]$. In the ``sgluon decoupling'' scenario, we estimate $\Delta \beta_\lambda(M_{O_S}) \simeq 24 R u_0$.
\subsection{Integrating out a Dirac gluino}\label{intgluino}
In the scenario where the gluino is much heavier than stops \cite{Fox:2002bu, Carena:2008rt, Muhlleitner:2008yw, Aebischer:2017aqa}, another EFT will be developed when the gluino is integrated out, and the leading two-loop effective potential can arise from the \textit{top-gluino-stop} bubble diagram, $V_{t\tilde{g}\tilde{t}}^{(2)}$. Meanwhile, removing the gluino from the theory breaks the $SU(3)_c$ supermultiplet explicitly, and different renormalization schemes should be used above and below mass thershold, i.e., using $\overline{\text{DR}}$ for $Q>M_3$ and $\overline{\text{MS}}$ while $Q < M_3$~\cite{Bagnaschi:2014rsa}. Thus, the \textit{top-gluon-top} diagram calculated in the aforementioned two renormalization schemes contribute to the threshold correction differently~\cite{Martin:2001vx}, and we denote the net contribution as ``$\lambda^{(2)}_{tgt, \text{reg}}$''. On the gluino mass threshold, the relevant effective potentials are represented as following,
\begin{figure}[h!]\label{fig:tfgst}
\centering
\includegraphics[width=5.5cm, height=3.0cm]{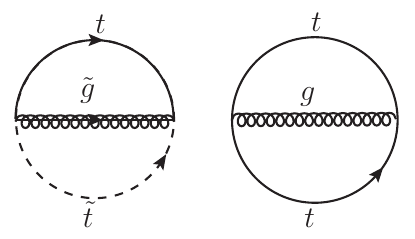}\;,\qquad
\includegraphics[width=2.5cm, height=3.0cm]{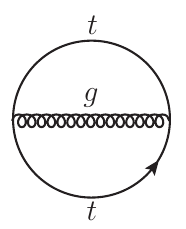}
\caption{$V_{t\tilde{g}\tilde{t}}^{(2)}$\;,\qquad$V_{tgt}^{(2)}(\overline{\text{DR}})$\;,\qquad\qquad $V_{tgt}^{(2)}(\overline{\text{MS}})$}\label{fig:tfgstntgt}
\end{figure}
and their corresponding fourth order derivative with respect to $H$ gives,
\bea
u_0^{-1}\lambda^{(2)}_{t\tilde{g}\tilde{t}, \text{der}}&=\frac{4}{r_f}+9+\frac{4}{3}\pi^2+6(\ln r_t+\ln r_f)+\left[-2+\frac{8}{3}\pi^2+36\ln r_f+4\left(1\right.\right.\\
&\left.\left.+2\ln r_f\right)\ln r_t\right]r_f+\mathcal{O}(r_f^2)\;,\quad\\
u_0^{-1}\lambda^{(2)}_{tgt, \text{reg}}=&-4(\ln r_t+2)\;,
\eea
where, $r_f=\mttilde^2/M_3^2,  r_t=m_t^2/M_3^2$. Like the discussion in the previous section, $\lambda_{\text{thr}}^{(2)}$ receives corrections from other one-loop diagrams. In addition to that, contributions from the $\overline{\text{DR}}-\overline{\text{MS}}$ conversion also need to be counted. The strategy for determining the connection between couplings in the $\overline{\text{MS}}$ and $\overline{\text{DR}}$ schemes is to relate each running parameter to physical observable which cannot depend on the choice of scheme.  Following this strategy, the complete dictionary for translating $\overline{{\text{MS}}}$ renormalized coupling to the $\overline{{\text{DR}}}$ couplings is derived by \cite{Martin:1993yx}. For instance, the Yukawa coupling $Y^{ijk}$ between a scalar $\phi_i$ and two chiral fermions $\psi_j, \psi_k$ is translated from one scheme to the other as, 
\beq
\left[Y_{\overline{\text{MS}}}^i\right]^{jk}=Y_{\overline{\text{DR}}}^{ijk}\left\{1-\frac{g^2\kappa}{2}[2C_2(r_i)-C_2(r_j)-C_2(r_k)]\right\}\;,
\eeq
where $C_2(r)$ is the quadratic Casimir invariant for a representation $r$. As a self-consistency check, we consider the $\mathcal{O}(g_t^4g_3^2)$ dependence of $\beta_\lambda$ calculated in both schemes. 
\begin{equation}
\beta^{\overline{\text{MS}}}_\lambda (g_t, g_3)= 12\kappa (\xi_{\tilde{t}}^2 - g_t^4)+64\kappa^2 g_3^2 \xi_{\tilde{t}}^2\;, \quad \beta^{\overline{\text{DR}}}_\lambda (g_t, g_3) = 0
\end{equation}
where $\beta^{\overline{\text{MS}}}_\lambda$ is computed from the relevant Feynman diagrams in $\overline{\text{MS}}$ scheme with the relation $\xi_{\tilde{t}}=g_t^2$ maintained at tree level, while $\beta^{\overline{\text{DR}}}_\lambda$ is obtained from $\beta^{\overline{\text{MS}}}_\lambda$ with the replacement $g_t \rightarrow g_{t, \overline{\text{DR}}}(1-\Delta g_{t, \text{reg}})$ and $\xi_{\tilde{t}}\rightarrow g_{t, \overline{\text{DR}}}^2$.
Since the Higgs quartic arises from electroweak D-term, the leading order $\overline{\text{DR}}$ result is independent of $g_3^2 g_t^4$ in the supersymmetric theory. 
At the gluino mass threshold, the one-loop corrections to $g_t$, $\xi_{\tilde{t}}$ and $m_{\tilde{t}}^2$ contribute to $\lambda_{\text{thr}}^{(2)}$ through both $\lambda_{\tilde{t}}^{1\ell}$
and the top quark box diagram $\lambda_t^{1\ell}=-6\kappa g_t^2\left[\lnbar(g_t^2|H|^2)+3/2\right]$ result in another two-loop Higgs quartic,
\bea\label{dlfg}
u_0^{-1}\lambda_{\text{thr}}^{(2)}(M_3)\simeq 4\bigg[\frac{\pi^2}{3}+2\bigg(\frac{\pi^2}{3}-1+3\ln r_f\bigg)r_f+\dots \bigg]\;, \quad (r_f\ll 1)\;.
\eea
Likewise, this threshold correction is finite when $r_t\rightarrow 0$ and $r_f\rightarrow0$. It is also remarkable to note that both the $\overline{\text{DR}}-\overline{\text{MS}}$ induced contributions form the effective potential and $\Delta g_{t, \text{reg}}$ cancelled compeletely against each other,
\beq
\lambda^{(2)}_{tgt, \text{reg}}+\frac{\partial \lambda_t^{1\ell}}{\partial g_t}g_t\Delta g_{t, \text{reg}}=0\;.
\eeq
The dominant $\beta$ function threshold correction is found to be $\Delta\beta_\lambda^{(2)}\simeq -24 u_0$ which will drive down the physical Higgs mass significantly if the stops are much lighter than the gluino. 
\subsection{Integrating out stops and all other SM superpartners}\label{st}
\begin{figure}[h!]
\centering
\includegraphics[width=0.45\textwidth]{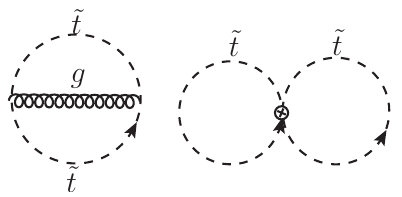}
\caption{$V^{(2)}_{\tilde{t}g\tilde{t}}$,\quad  $V^{(2)}_{\tilde{t}\tilde{t}}$.}
\label{fig:stopdiag}
\end{figure}

The stop mass is the last threshold beyond standard model, but still well above EW scale. Repeating the analysis in the previous sections via the relevent diagrams in Fig~\ref{fig:stopdiag}, we get $\lambda_{\text{thr}}^{(2)}(\mttilde)=-u_0$.  

Wherein, the two bubble diagrams lead  $\lambda_{\tilde{t}g\tilde{t}, \text{der}}^{(2)}=-u_0$. 
In order to work out the Higgs mass, we also need to include the one-loop contributions to $\lambda$ from other SM superpartners $\Delta\lambda^{1\ell}$ as well as the SM RGEs. In the ``gaugeless'' and ``minimal mixing'' limit, the contributions to $\Delta\lambda^{1\ell}$ are those proprotional to the fourth power of a third-family Yukawa coupling,
\beq
\Delta\lambda^{g_f^4}=2\kappa\sum_{f=t, b,\tau}g_f^4 N_c^f\lnbar(m_{\tilde{f}_L} m_{\tilde{f}_R})\;,
\eeq
where $\lnbar(x)=\ln(x/Q^2)$ and the SM-like Yukawa couplings relates to their MSSM counterparts for each fermion species $f$ via $g_f=y_f\sin\phi_f$ with $\phi_t=\beta$, $\phi_b=\phi_\tau=\pi/2-\beta$; $N_c^f$ is the number of colors, and $m_{\tilde{f}_i}$ are the soft SUSY-breaking sfermion masses\cite{Bagnaschi:2017xid}. Although $y_b$ and $y_\tau$ could be potentially large when $\tan\beta$ increases, the top coupling $g_t$ still dominates the radiative correction since the down-type Yukawa contributions are supressed by the factor of $\cos\beta$. 
The full result for one-loop threshold corrections can be found in \cite{Bagnaschi:2014rsa} which neglects all Yukawa couplings except $g_t$, and \cite{Bagnaschi:2017xid} which additionally computed the full dependence on the bottom and tau Yukawa couplings. To obtain the 2-loop $\beta$-function for $\lambda_R$, one may either consult the results provided by Luo, Wang and Xiao\cite{Luo:2002ey, Luo:2002ti}, which refined and improved the seminal works of Machacek and Vaughn\cite{Machacek:1983tz, Machacek:1983fi, Machacek:1984zw} or use the public code PyR@TE\cite{Lyonnet:2013dna, Lyonnet:2016xiz} which performs the calculation in $\overline{{\text{MS}}}$ renormalization scheme. As an alternative choice, taking the model parameters in the full theory as inputs, the public Mathematica package SUSYHD~\cite{Vega:2015fna} provides an automatic calculation of Higgs boson pole mass in the SM according to the analytic formulae in~\cite{Bagnaschi:2017xid}. In the next section we will present our numerical analysis based on SUSYHD.
\section{Numerical results}\label{num}
To consider the quantitative effects of the presence of Dirac gauginos, we considered two basic scenarios. In the first, we focus on a SUSY model with only a Dirac gluino extended beyond the MSSM. This provides a good sense of the size of effects that can arise. However, other scenarios are of great interest. In particular \cite{Fox:2014moa} considered a scenario where the Dirac gauginos were at an intermediate scale, pulling all the scalars up as well. Higgsinos can remain light providing a dark matter candidate, and the Higgs is light from tuning.  \cite{Fox:2014moa} considered two variants - ``pure Dirac'' and ``hypercharge impure'' - which we will study as well.

We implement numerical calculation which focuses on matching and running of the renormalized Higgs quartic coupling as mentioned in sec.~(\ref{higgsmassintro}). The main purpose here is to illustrate the dependence of the quantum corrections to $m_h$ on various relevant parameters. Above $\mttilde$, we use the analytic expressions of both threshold corrections and $\beta$ functions calculated in the previous section to do estimation up to the order of $\mathcal{O}(g_3^2 g_t^4)$ in Dirac gluino model and $\mathcal{O}(g_t^4g_2^2+g_2^6)$ in pure Dirac and hypercharge impure model respectively. Below that scale where the SUSY is completely broken, we will use SUSYHD to complete the computation which includes one-loop threshold corrections from SM superpartners at $\mttilde$, 2-loop SM radiative corrections $\delta\lambda(m_t)$ at the electroweak scale.  As to the SM parameters through out the calculation, we will adopt the pole mass of the top quark $m_t=173.34$ GeV\cite{ATLAS:2014wva}, the $Z$ boson mass $m_Z=91.187\gev$, $m_W=80.385\gev$, the Fermi constant $G_F=1.16637\times 10^{-5} \text{GeV}^{-2}$, the strong gauge coupling in the $\overline{\text{MS}}$ renormalization scheme $\alpha_s(m_Z)=0.119$\cite{Tanabashi:2018oca}.

\subsection{Dirac Gluino Model}
\begin{figure}[H]
\centering
\includegraphics[width=0.45\textwidth]{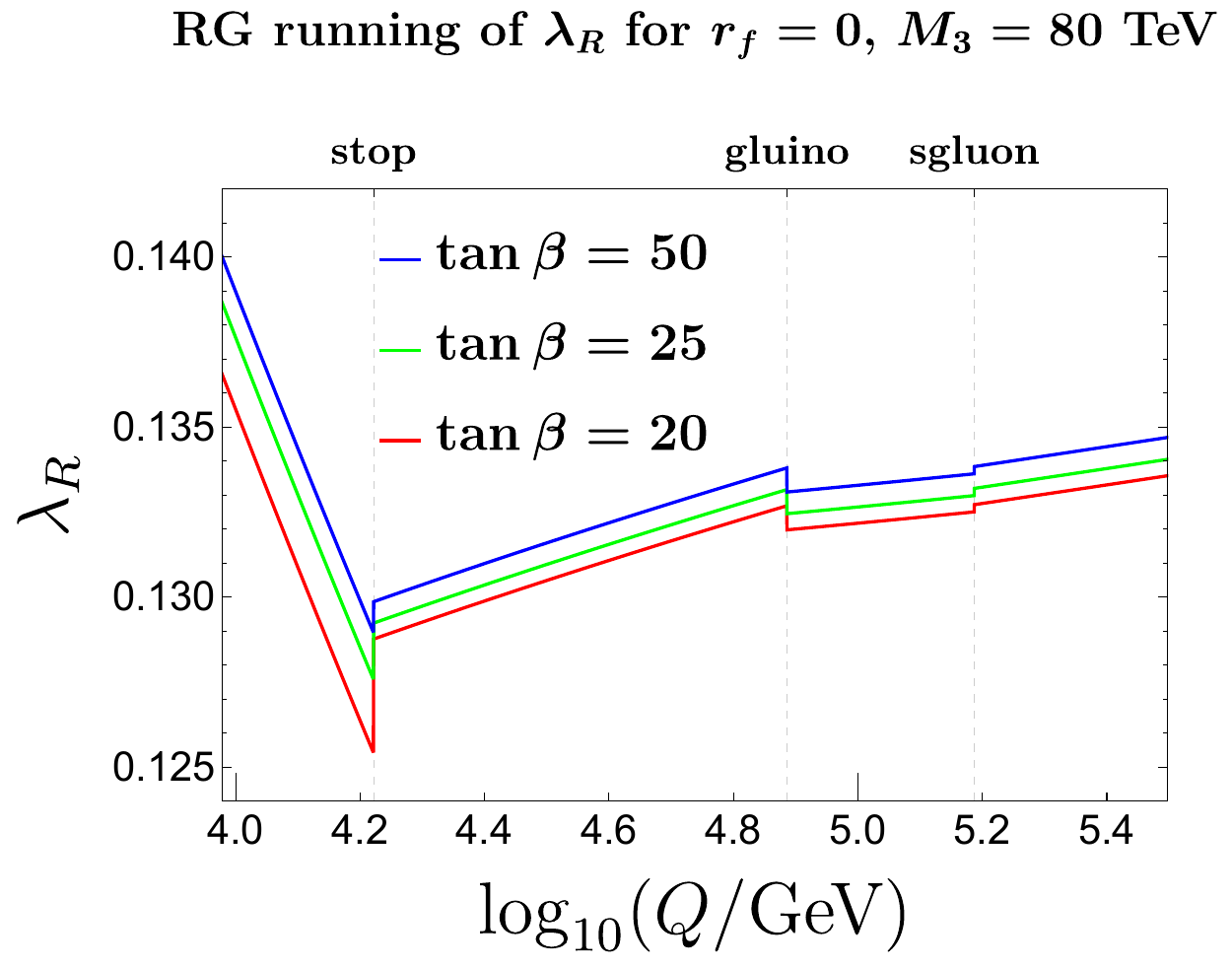}
\includegraphics[width=0.45\textwidth]{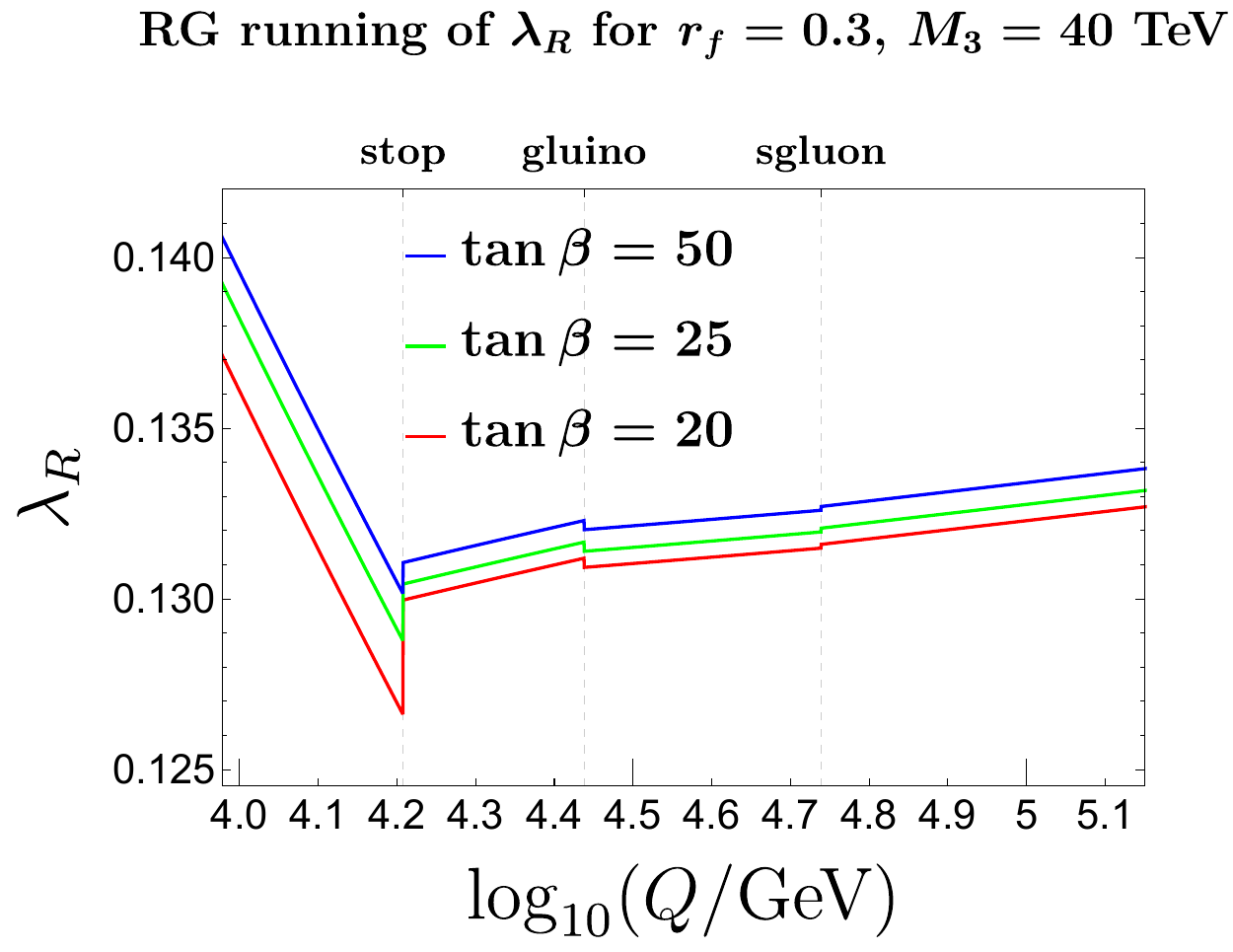}
\caption{The running of $\lambda_R$ for different $r_f$ and other model parameters. For $\tan\beta=20$, we take the effective common SUSY-breaking assumption. For $\tan\beta=25$ and $\tan\beta=50$ we choose the soft masses $(m_A, m_{12})$ equal to $(2\mttilde,\; 0.5\mttilde)$ and $(1.5\mttilde, 0.3\mttilde)$ respectively. }\label{fig: runlambda}
\end{figure}

Fig.~(\ref{fig: runlambda}) represents the trajectories of the Higgs quartic coupling $\lambda_R$ running according to renormalization scale $Q$. In the absent of soft sgluon mass, i.e. $R=1/4$, we provide the numerical results for different values of $\mttilde$ and $M_3$. Besides that, we can also size the heavy-Higgs mass $m_A$, the degenerate wino and bino mass $M_1=M_2=m_{12}$ and $\tan\beta$ to generate other parallel trajectories. We find that the amount of the threshold correction increases with increasing $m_A$ or decreasing $m_{12}$. Being an input parameter the high scale soft-breaking stop mass is chosen as $\mttilde^2=0$ and $\mttilde^2=0.3 M_3^2$ for the left and right figures respectively.

On the left panel, we set $M_3=82.9\tev$ so that the resulting stop mass ($\sim 16.6\tev$) can reproduce the realistic Higgs quartic coupling in the absence of squark mixing. 
  The red line in Fig.~(\ref{fig: runlambda}) depicts the effective common SUSY-breaking scale with $\tan\beta=20$, in which $m_A=m_{12}=\mttilde\simeq 0.217 M_3$. As one can read from E.q.~(\ref{dlsg}) and E.q.~(\ref{dlfg}) the amount of radivative correction from matching boundaries is insensitive to the sgluon and the gluino masses. Correspondingly, the two-loop threshold corrections received by the mass of Higgs boson are rather limited. Applying e.q.~(\ref{mhpole}) we  find that $\delta m_h^{\rm pole}\simeq-33\mev$ from the sgluon and $\delta m_h^{\rm pole}\simeq 170\mev$ from the gluino. 
  
  For Dirac-gluinos heavier than the stops, $\lambda_R$ runs significantly between the two mass thresholds. It is this running effect that provides the dominant $\mathcal{O}(g_3^2 g_t^4)$ contribution to the Higgs quartic and influences $m_h$ as much as $-0.61$ GeV. At $Q=\mttilde$, the one-loop threshold correction from superpartners are negative and contribute $-1.39\gev$ to Higgs pole mass. For comparison, keeping $\mttilde$ fixed and adjusting other model parameters we can see the variation of $\lambda_R$.  
The green line represents the model parameters fixed at $m_A=2 \mttilde$, $m_{12}=0.5\mttilde$, $\tan\beta=25$. From the plot, we find that the tree level Higgs quartic coupling is lifted up and stop mass threshold correction is suppressed which results in a $0.41\gev$ increase in the physical Higgs boson mass. The similar curve can be obtained by assigning $m_A=1.5 \mttilde$, $m_{12}=0.3\mttilde$, $\tan\beta=50$ which is portrayed by the blue line. $m_h$ obtained in this scenario is $0.66\gev$ heavier than that calculated under the effective common SUSY-breaking assumption. On the right panel, allowing a nonzero tree-level soft stop mass, the gluino mass we need for providing the correct value of Higgs quartic decreases to $40\tev$.  The reason is twofold: firstly, one needs less of a radiative correction from the gluino to generate the desired stop pole mass; and secondly, the soft stop mass shortens the interval between gluino and the stop masses in which the running effect of $\lambda_R$ makes a negative contribution to the low scale Higgs quartic.

\begin{figure}[H]
\centering
\includegraphics[width=0.45\textwidth]{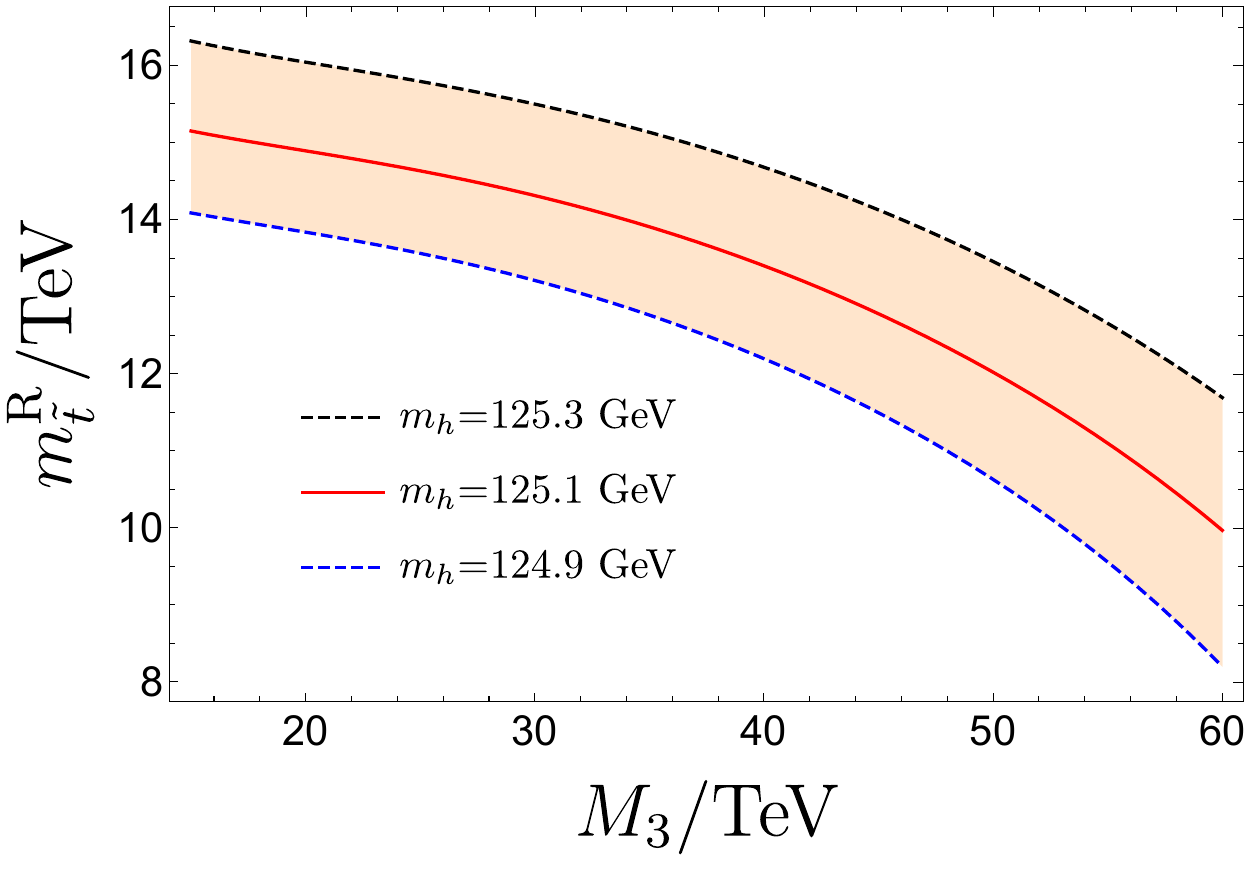}
\includegraphics[width=0.45\textwidth]{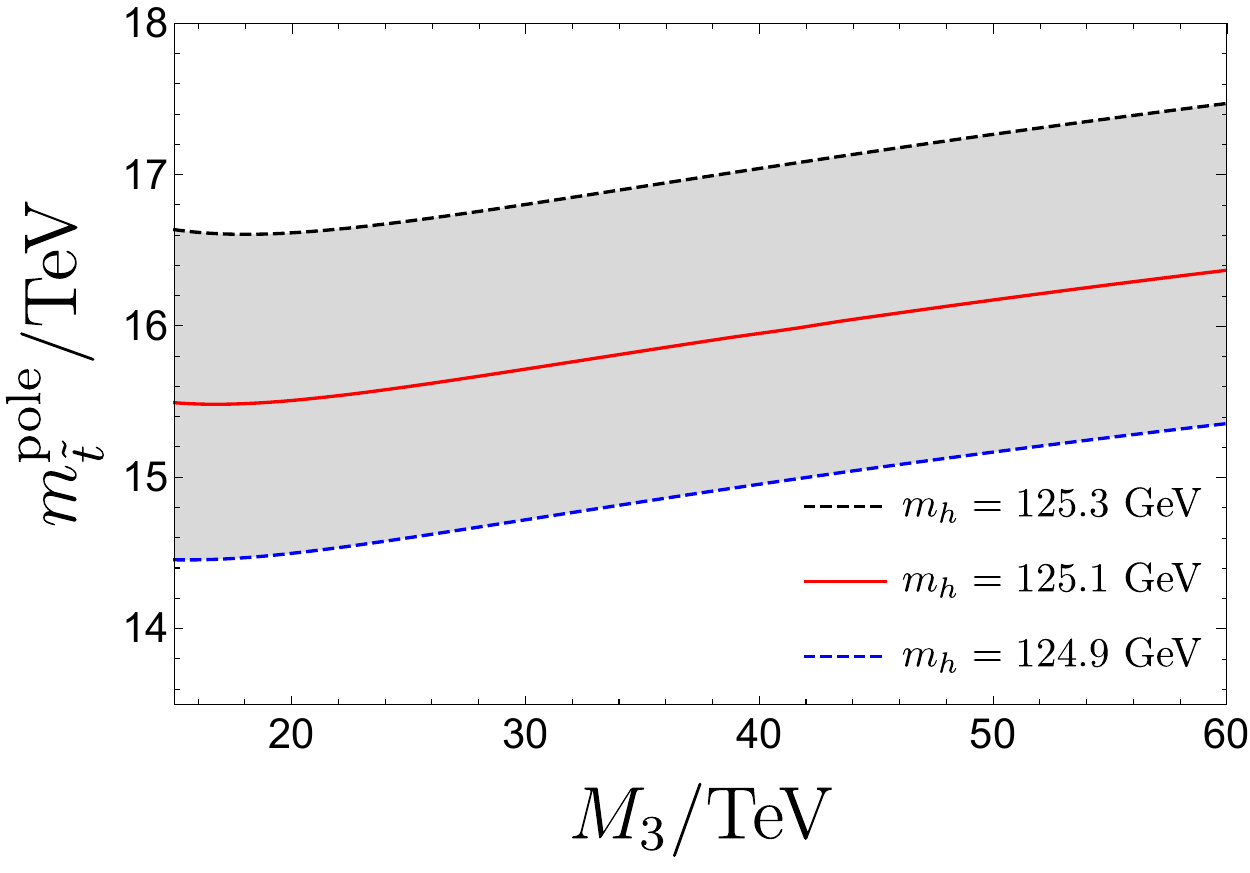}
\caption{For $m_h=125.09\pm0.24$ GeV, the relation between the gluino mass and the soft-breaking stop mass. The solid red lines represent the mean value of Higgs boson mass, the black (blue) dashed lines represent the upper (lower) limits of the errors, and the shaded areas mark the regions where the model parameters are consistent with the measurement.  }\label{fig:m3mst}
\end{figure}

We can also vary both stop mass and the gluino mass in such a way that the resulting Higgs pole masses satisfy the experimental constraints; by ``stop mass'' we mean both the renormalized soft-breaking stop mass $\mttilde^R$ and the pole mass of stop $\mttilde^{\text{pole}}$. The measurement of the Higgs boson mass is performed by ATLAS for both $H\rightarrow\gamma\gamma$ and $H\rightarrow Z Z\rightarrow 4 \ell$ channels\cite{Aaboud:2018wps} and by CMS experiments \cite{Sirunyan:2017exp} for $4\ell$ channel; including the individual and the combined measurements, its experimental value is taken to be $m_h=125.09\pm0.24$ GeV\cite{Aad:2015zhl}. In Fig.~(\ref{fig:m3mst}) we plot the resulting Higgs boson mass in the effective common SUSY-breaking scenario with $m_A=m_{12}=\mttilde^{\text{pole}}$ and $\tan\beta=20$.  

The contours in the left panel are plotted in the $M_3-\mttilde^R$ plane, from which we can tell both $M_3$ and $\mttilde^R$ provide positive contributions to the pole mass of Higgs boson. The curves on the right panel are represented in the $M_3-\mttilde^{\text{pole}}$ plane which imply that the smaller  Dirac-gluino mass we put in the model, the smaller stop pole mass is needed to reproduce the physical Higgs mass.
\subsection{Pure and Hypercharge Impure Model}
Observing that the Higgs quartic coupling runs to zero at an intermediate scale, a framework is proposed for split SUSY models~\cite{Fox:2014moa}, in which gauginos acquire intermediate scale Dirac masses $(\sim 10^{8-12})$ GeV and the higgsino mass settles near the weak scale due to the approximate $U(1)_{PQ}$ symmetry. At this scale, the weak gauge coupling is approximately equal to the strong gauge coupling and therefore their radiative corrections are comparable to that of the strong one.
\begin{figure}[H]
\centering
\includegraphics[width=0.45\textwidth]{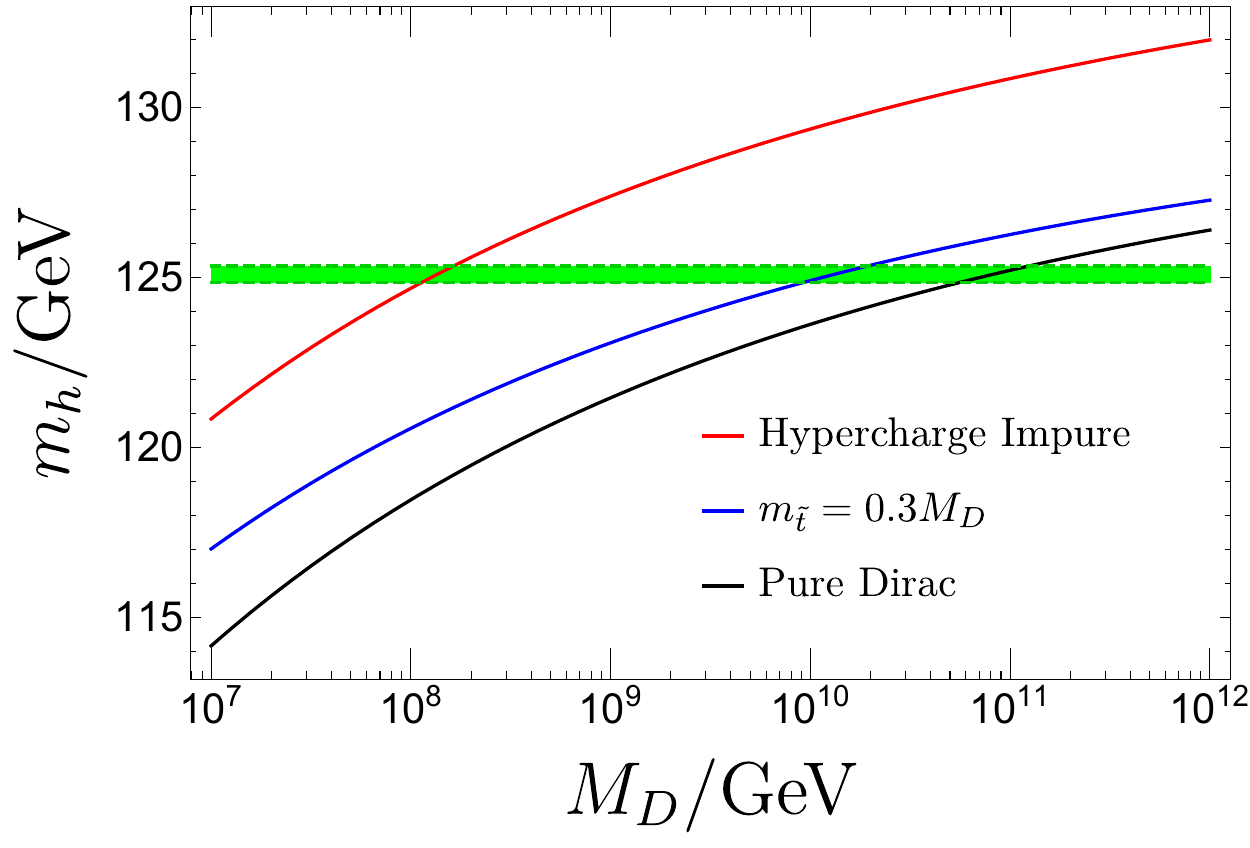}
\includegraphics[width=0.49\textwidth]{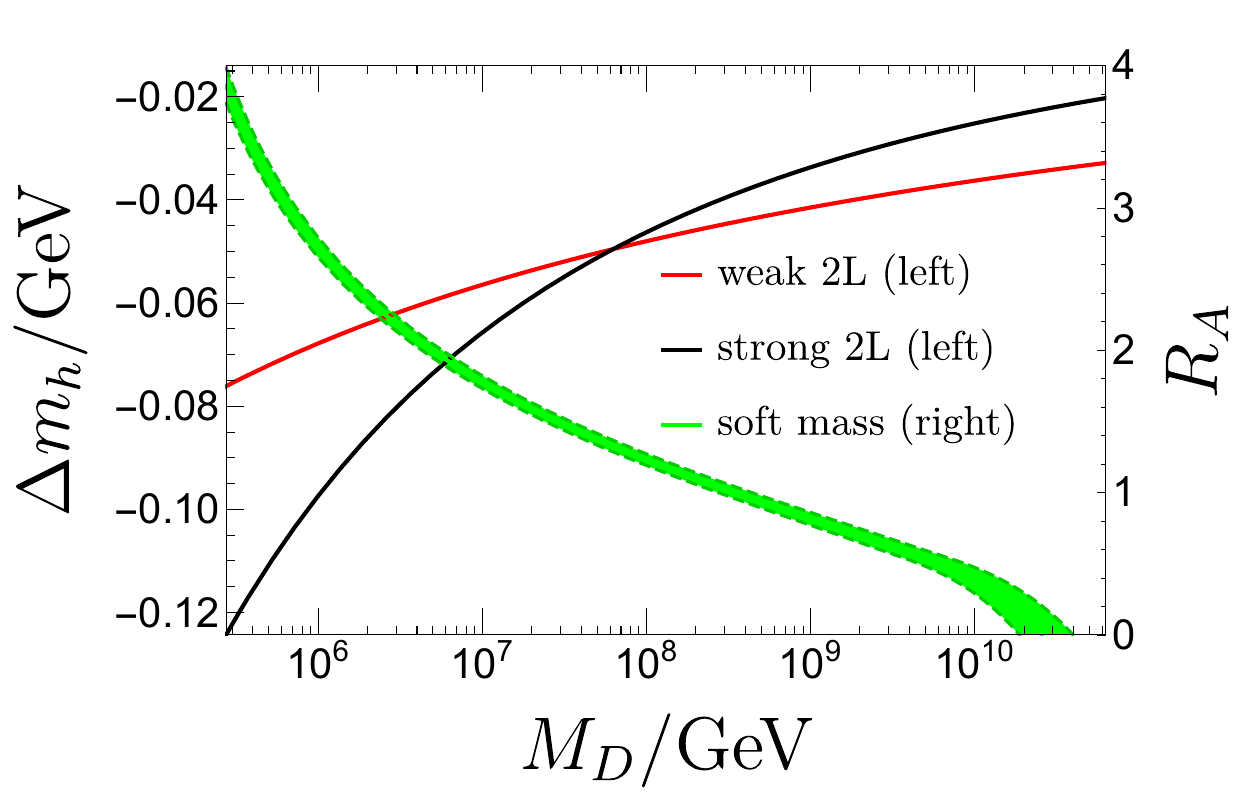}
\caption{The left panel shows the predicted Higgs mass $m_h$ in split supersymmetry as a function of the universal Dirac gaugino mass $M_D$ in hypercharge impure model (red line), pure Dirac model with and without tree-level Dirac mass (blue line and black line, respectively). We have chosen $\mu=1\tev$, $\tan\beta=10$ and shaded the experimentally favored range $m_h =125.09\pm 0.24\gev$ in green. The plot on the right presents the two-loop correction to the Higgs quartic from strong interactions and weak interactions in pure gaugino model which are depicted by the black line and the red line respectively. The green band shows the soft adjoint scalar mass one needs to give rise to the necessary size of Higgs quartic in the pure Dirac model.}
\label{fig:splitMDmh}
\end{figure}

On the left panel of Fig.~(\ref{fig:splitMDmh}), we present the variation of the Higgs pole mass versus the mass of Dirac gaugino up to $\mathcal{O}(g_3^2g_t^4+g_2^2 g_t^4+g_2^6)$. Without a tree-level Higgs quartic coupling, the gaugino mass is required to be around $ 10^{11}\gev$ in the pure Dirac model so that a large enough stop mass can be produced to generate the needed Higgs quartic. To generate sufficient $\lambda_R$, we can add a nonzero soft stop mass $m_{\tilde{t}}$ in the tree level Lagrangian. In the pure Dirac model, for $m_{\tilde{t}}=0.3 M_D$ we find $\sim 10^{10}\gev$ gaugino mass is needed. Alternatively, in the so-called ``hypercharge impure'' model, the bino does \textit{not} acquire a Dirac mass, and therefore tree-level Higgs quartic coupling $\lambda_R(m_{\tilde{t}})=3g_1^2\cos^2(2\beta)/20$ is re-introduced. For $\tan\beta= 10$ in our example, the desired gaugino mass is reduced to $\sim 10^8\gev$. On the right panel the red and the black lines refer to the two-loop impacts on the Higgs mass from the weak interaction ($g_2^2 g_t^4+g_2^6$) and the strong interaction $g_3^2g_t^4$ respectively, which can be read in the left vertical axis. As one may anticipate, the Higgs quartic receives equal amount of radiative corrections from the strong and the weak interactions around the intermediate scale. The green band in the same panel indicates the value of the tree-level adjoint scalar mass $m_{\Sigma_A}$ which brings about the correct $m_h$, where we have defined $R_A=m_{\Sigma_A}/M_D$ at the right vertical axis to measure it.
\section{Conclusion}\label{conclusion}
In this work, we revisited the leading order two-loop corrections on the Higgs boson mass in the Dirac-gaugino model. According to \cite{Diessner:2015yna}, the SUSY model with Dirac gaugino naturally provides positive impact of two-loop corrections on the mass of the Higgs boson. Comparing with MSSM, the shift of Higgs boson mass is typically +5 GeV and further increased with large Dirac gluino mass. However,  \cite{Braathen:2016mmb}  performed an explicit two-loop calculation pointed out that this significant two-loop correction was unphysical and should be attributed to the straightforward implementation of $\overline{\text{DR}}$ renormalization scheme in automated calculations. They demonstated that the actual correction from Dirac gaugino is rather limited when the $\overline{\text{OS}}$ scheme is adopted. 
Both of their works are performed in the so-called ``fixed-order'' approach in the sense that the RG running of model parameters is ignored and turns out to be a suitable choice for low-scale SUSY prediction. 

We presented a similar calculation however in a different context in this paper, where SUSY scales are split and live well above the electroweak scale. In our setup, the ``fixed order" formalism becomes inaccurate for one needs to resum large logarithms of the scale ratios. 
Therefore, we adopted the so-called ``match-and-run'' procedure which is based on the successive decoupling of particles at the scale of their masses. $\lambda_R$ runs from one mass scale to another according to its $\beta$ function and, the effective-potential technique is only applied while computing the two-loop corrections to Higgs quartic couplings around the mass thresholds. In the high scale theory, we renormalized the model parameters in the SUSY-preserving $\overline{\text{DR}}$ scheme. However, to allow for the direct implementation of existing SM codes we chose the $\overline{\text{MS}}$ scheme for the regime below the gaugino mass. The threshold correction is obtained by comparing the scattering amplitudes $\mathcal{M}$ calculated in both high and low scale EFTs. We emphasized that although $\mathcal{M}$ can blow up unphysically in both sides, the divergent components always cancel out and there remains a finite   correction on the mass threshold. The loop correction to the pole mass of the Higgs boson is independent of renormalization scheme as stated in Ref.~\cite{Braathen:2016mmb}. Numerically, the two-loop effects are found to be fairly modest -- up to $\mathcal{O}(g_3^2g_t^4+g_2^2 g_t^4+g_2^6)$, the shifts of Higgs boson mass are typically below one GeV in the \textit{supersoft} limit. 

The absence of any signal of new physics at the LHC requires us to examine closely all of our models of the weak scale. The role of SUSY in BSM physics and as an explanation of the hierarchy problem is especially unclear. Given its fairly narrow predictions of the mass of the Higgs boson, it is worth studying carefully these quantitative requirements. While the effects we have found are small, they give us confidence in interpreting the appropriate ranges of parameters in a variety of BSM models.   
\vskip 0.25in

\noindent{\bf Note added:} As this work was in preparation, \cite{Braathen:2016mmb} appeared which contains the cancellation of unphysical two-loop corrections. I had achieved the same basic result in sec.~\ref{matchrun} but had not finished the paper when they put theirs out.
\vspace{.3cm}
\acknowledgments
I am indebted to Prof. Neal Weiner for discussions and guidance in this project. I am also grateful to Johannes Braathen for his insightful comments and careful reading of the manuscript. This project is supported in part by the National Science Foundation under grant PHY-1620727. 
\appendix
\section{Expressions of threshold corrections}\label{appendix}
In this appendix, we present the two-loop correction to the Higgs quartic coupling. When evaluating two-loop effective potential like $F_{SSS}(x, y, z)$ or $F_{FFS}(x,y, z)$, one always encountours the loop integral $I(x, y, z)$ which is first introduced in \cite{Ford:1992pn, Kotikov:1991hm}. A more explicit form of such integral can be found e.g. in eqs. (D1)-(D3) of \cite{Degrassi:2009yq}, wherein a particular function $\Phi(x, y, z)=\Phi(x/z, y/z, 1)$ and its recursive relation for derivatives is defined which proves very useful for obtaining compact analytical results. To obtain approximate expression we found it is most convenient to do series expension w.r.t. $u=x/z$ and $v=y/z$.
For exmaple, in the case where there exist hierachy between variables : $y\ll x<z$, it would be practical to have,
\beq\label{phi}
\Phi(u,v,1)=\frac{1}{1-u}\left[\ln u\ln v-2\ln(1-u)\ln u-2 \text{Li}_2(u)+\frac{\pi^2}{3}\right]+\mathcal{O}(v)\;.
\eeq
\subsection{Two-Loop Contribution From $SU(3)_c$ }\label{fullhiggsqtc}
The strong two-loop contributes to Higgs quartic through derivatives of effective potential. We can express stop mass as a function of field-dependent top mass $m_t=g_t|H|$ and re-write e.q.~(\ref{vder}) as~\cite{Bagnaschi:2014rsa} 
\begin{equation}
\Delta\lambda^{(2)}_{\text{der}}=\left.\frac{g_t^4}{2}(2\mathcal{D}+4 m_t^2\mathcal{D}_3+m_t^4\mathcal{D}_4)V^{\text{eff}}\right\vert_{m_t\rightarrow 0}\;,\quad \mathcal{D}_i=\left(\frac{d}{dm_t^2}\right)^i
\end{equation} 
The explicit expressions for diagram Fig~(\ref{fig:stostnstst}) and Fig~(\ref{fig:stopdiag}) are given as following,
\bea
u_0^{-1}\lambda^{(2)}_{\tilde{t}O^a\tilde{t}, \text{der}}(Q)&=-\frac{4(1-\lnbar M_{O_s}^2)}{r_f}-\frac{4R}{(1-4r_s)^{3/2}}\bigg[\frac{\pi^2}{3}+4\sqrt{1-4r_s}\ln r_s-\ln^2r_s+2\ln(x_s^-)\\
&-4\text{Li}_2(x_s^-)\bigg]
\eea
\bea
u_0^{-1}\lambda^{(2)}_{t\tilde{g}\tilde{t}, \text{der}}(Q)&=\frac{4}{r_f}\left[\frac{1}{(1-r_f)^2}-\lnbar M_3^2\right]+2(1-2\lnbar\mttilde^2)\lnbar M_3^2-4(2+\ln r_t)\lnbar\mttilde^2\\
&+\frac{1}{(1-r_f)^2}\bigg[1+\frac{4}{3}\pi^2-(8-3r_f)r_f+2(3-4r_f+r_f^2+2\ln r_f)\lnbar r_t\\
&+14\ln r_f-4\ln r_f\ln(1-r_f)-8\text{Li}_2(r_f)\bigg]
\eea
where $\text{Li}_2(z)=-\int_0^z dt [\ln(1-t)/t]$ is the dilogarithm function and $x_s^-=(1-\sqrt{1-4 r_s})/2$.
\bea
u_0^{-1}\lambda^{(2)}_{\tilde{t}g\tilde{t}, \text{der}}(Q)=-1+5\lnbar\mttilde^2-3\lnbar^2\mttilde^2\;,\quad u_0^{-1}\lambda^{(2)}_{\tilde{t}\tilde{t}, \text{der}}(Q)=-1+\lnbar\mttilde^2+\lnbar^2\mttilde^2\;.
\eea
\bea
u_0^{-1}\lambda^{(2)\overline{\text{DR}}}_{t g t, \text{der}}(Q)=1+16\lnbar m_t^2+6\lnbar^2m_t^2\;,\quad u_0^{-1}\lambda^{(2)\overline{\text{MS}}}_{t g t, \text{der}}(Q)=9+20\lnbar m_t^2+6\lnbar^2m_t^2\;,
\eea
Here we explicitly reserved the terms that contains $\lnbar m_t^2$ in the last two equations. As discussed in sce.~\ref{intgluino}, these terms are cancelled out on the matching boundary when one-loop factors are taken into account. 
\subsection{Two-Loop Contribution Form $SU(2)_L$ Sector}
\begin{figure}[h!]
\centering
\includegraphics[width=18.0cm, height=14.0cm]{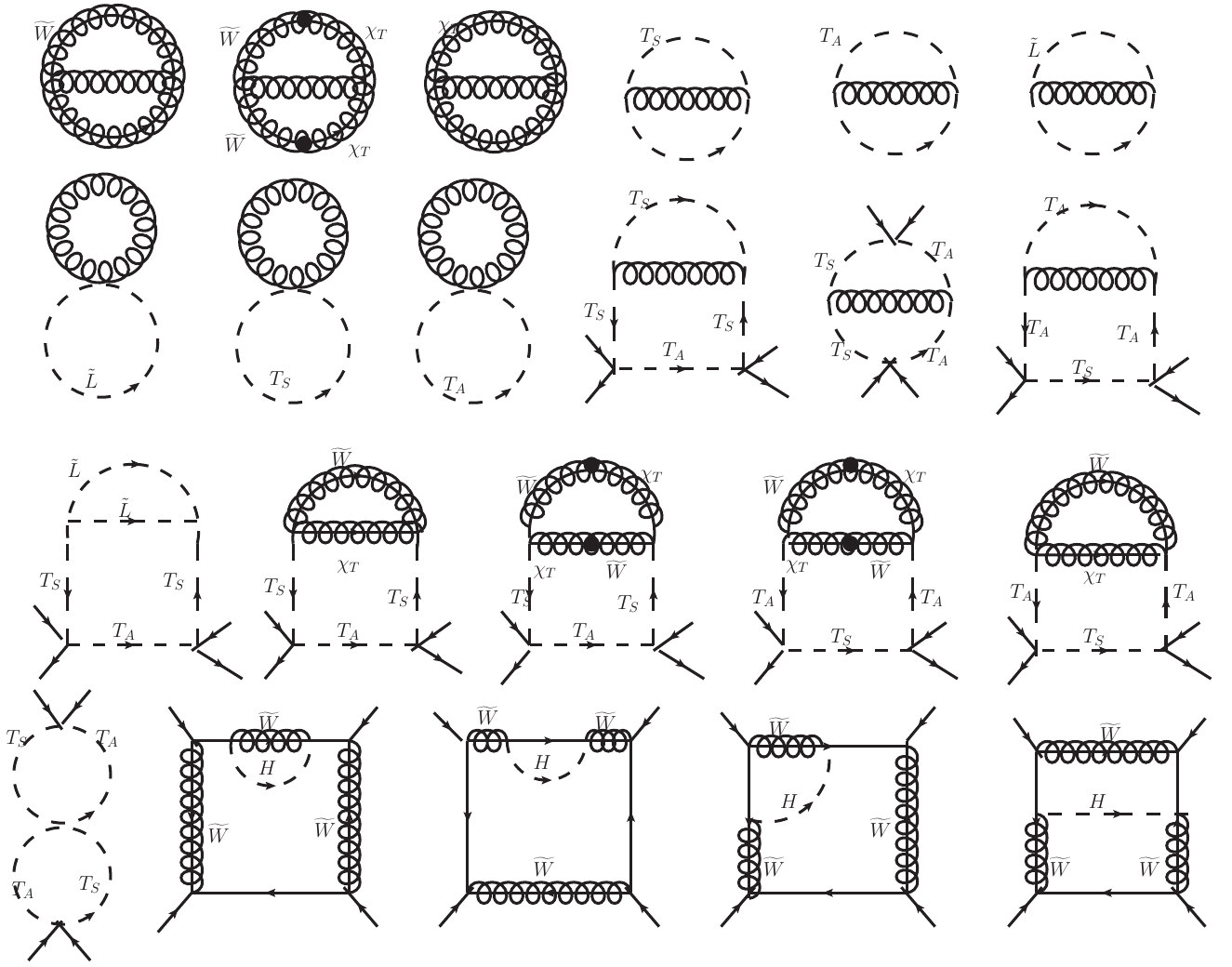}
\caption{\small{Feynmans diagram for weak two-loop contributions, where $\chi_T$ is the Dirac partner of the wino $\widetilde{W}$, and $T_S(T_A)$, are the $SU(2)_L$ triplets (pseudo)scalar, a.k.a, swino.}}
\label{fig:2lw}
\end{figure}

In the high energy regime where the electroweak symmetry is restored, two-loop corrections from top and $SU(2)_L$ gauge supermultiplets. Replacing $u_0$ with $u_2=9g_t^4 g_2^2/2$, it is straightforward to get the radiative correction from the weak interaction of top/stop. We list the additional  Feynman diagrams of the effective potential in Fig.~(\ref{fig:2lw}),

As one may be aware that most of electroweak two-loop contributions resort to Feynman diagram approach~\cite{Martin:2003it, Goodsell:2015ira}. Having analytical experessions for the bubble diagrams, the corresponding Higgs quartic diagrams with zero external momenta can be obtained via the difference quotients of the effective potential w.r.t. particular variables, e.g.
\begin{equation}
 F_\alpha^{(1,0,0)}(x, u; y, z)\equiv \frac{F_\alpha(x, y, z)-F_\alpha(u, y, z)}{x-u} 
\end{equation}
for $u=x$ the difference quotient becomes the partial derivative, i.e. $F_\alpha^{(1,0,0)}(x, x; y, z)=\partial_x F_\alpha(x, y, z)\equiv F_\alpha (x', y, z)$. We can classify the radiative corrections to $\lambda_R$ in the following cases.

\subsubsection{Contributions from $FFV$, $\overline{FF}V$, $SSV$ and $SV$ diagrams}

Since $SU(2)_L$ gauge vectors have Higgs field-dependent squared-mass $W\equiv m_W^2=g_2^2|H|^2/2$, the radiative corrections are arised from vertices on the vector line in this case. The bubble-diagrams corresponds to $V(W)$ gives, 
\bea
V(W)&=6g_2^2\left[F_{FFV}(\wtilde, \wtilde, W)-\wtilde F_{\overline{FF}V}(\wtilde, \wtilde, W)\right]+\frac{3g_2^2}{2}\left[ F_{SSV}(T_S,T_S,W)\right.\\
&\left.+2F_{SV}(T_S, W)+(T_S\leftrightarrow T_A)\right]+\frac{3 g_2^2}{4}n_f\left[F_{SSV}(\tilde{L},\tilde{L},W)+2F_{SV}(\tilde{L}, W)\right]\;,
\eea
 and the effective potential induces the Higgs quartic via, 
\begin{equation}
\Delta\lambda_W=\left.\frac{g_2^4V(W'')}{4}\right\vert_{W\rightarrow 0}\;,
\end{equation}

where $n_f=12$, $\wtilde$ is the wino mass and $T_{S, A}$ refers to the masses of scalar triplet and the pseudoscalar triplet respectively. 
\subsubsection{Contributions from $SSV$, $SS$, $SSS$ and $FFS$ diagrams}
Due to the gauge interactions between $H$ and the $SU(2)_L$ adjoint scalars $T_S$ and $T_A$, quartic contributions in this case are induced from the vertices on the scalar lines.
If the two Higgs vertices are attached on the same scalar line, the radiative correction derived from this catalog will take the form of, 
\begin{equation}
\Delta\lambda_{S_1} =\frac{g_2^4}{6}V^{(2,0)}(x,z,x; y)\;,
\end{equation}
As a consequence, in the Dirac wino model we have, 
\bea
\Delta\lambda_{S_1}&=\frac{g_2^6}{4}F_{SSV}^{(2,0,0)}(T_S, T_A, T_S; T_S, 0)+(T_A\leftrightarrow T_S)+g_2^6 F_{FFS}^{(0,0,2)}(\wtilde, \widetilde{W}; T_S, T_A, T_S)\\
&+g_2^6\wtilde F_{\overline{FF}S}^{(0,0,2)}(\widetilde{W}, \widetilde{W}; T_S, T_A, T_S)+g_2^6 F_{FFS}^{(0,0,2)}(\wtilde, \widetilde{W}; T_A, T_S, T_A)\\
&-g_2^6\wtilde F_{\overline{FF}S}^{(0,0,2)}(\widetilde{W}, \widetilde{W}; T_A, T_S, T_A)+\frac{g_2^6}{2} n_f M_D^2 F_{SSS}^{(0,0,2)}(\tilde{L}, \tilde{L}; T_S, T_A, T_S) 
\eea

If the Higgs vertices are on two defferent scalar lines one finds the ``sunrise'' diagram gives, 
\begin{equation}
\Delta\lambda_{S_2}=\frac{g_2^4}{3}V^{(1,1)}(x,y)=\frac{1}{8} g_2^6 F_{SSV}^{(1,1,0)}(T_S, T_A; T_S, T_A; 0)
\end{equation}
and the ``snowman'' diagram is evaluated as,
\begin{equation}
\Delta\lambda_{S_3}=\frac{3g_2^6}{2} B^2(T_S, T_A)\;,\quad B(x, y)=\frac{x(\lnbar x-1)-y(\lnbar-1)}{x-y}
\end{equation}
There is another ``snoman'' contribution which comes from the two vertices attaching on the same loop,
\begin{equation}
\Delta\lambda_{S_4}=\frac{g^2 N_c(N_c-1)\kappa^2}{2}\frac{(r-1-\ln r)(\lnbar T_S -1)}{(r-1)^2}\;,\quad r=\frac{T_A}{T_S}
\end{equation}
However, this contribution is cancelled by its one loop conterpart and does not show up in the final result. 

\subsubsection{Contribution from Higgs-Higgsino-wino box}
The Higgs quartic corrections in the last case are originated from the Higgs-Higgsino-wino vertex. By the direct computation of the last four Feynman diagrams in Fig~(\ref{fig:2lw}), the weak two-loop is given by, 
\bea
4\Delta\lambda_{F}&=5 g_2^6 f_{FFS}(\wtilde'',0,0)+15 g_2^6f^{(2, 0, 0)}_{FFS}(\wtilde, 0,\wtilde;\wtilde,0)\\
&+20 g_2^6f_{SSS}(\wtilde', \wtilde, 0)+7g_2^6f_{FFS}^{(1, 1, 0)}(\wtilde, \wtilde;\wtilde, 0; 0)
\eea


\end{document}